\documentclass[11pt,english]{article}
\usepackage{times}
\usepackage[T1]{fontenc}
\usepackage{geometry}
\usepackage{subfigure}
\usepackage{amsmath}
\usepackage{graphicx}
\usepackage{setspace}
\usepackage[numbers]{natbib}

\makeatletter

\providecommand{\tabularnewline}{\\}

\newcommand{\lyxaddress}[1]{
\par {\raggedright #1
\vspace{1.4em}
\noindent\par}
}

\date{}

\usepackage{babel}
\makeatother
\begin{document}

\title{Nanostructure and microstructure of laser-interference induced dynamic patterning of Co on Si}

\author{L.Longstreth-Spoor, J.Trice, {*}H.Garcia, C.Zhang, and Ramki Kalyanaraman%
\thanks{ramkik@wuphys.wustl.edu%
} }

\maketitle

\lyxaddress{\begin{center}Department of Physics, Washington University in St. Louis, MO 63130\par\end{center}}

\lyxaddress{\begin{center}Center for Materials Innovation, Washington University in St. Louis, Mo 63130\par\end{center}}

\lyxaddress{\begin{center}{*}Department of Physics, Southern Illinois University, Edwardsville, IL 62026\par\end{center}}

\begin{abstract}
We have investigated the nanostructure and microstructure resulting from ns laser irradiation
simultaneous with deposition of Co films on Si(001) substrates. The spatial order and length
scales of the resulting nanopatterns and their crystalline microstructure were investigated as
a function of film thickness $h$ and laser energy density $E$ using a combination of atomic
force, scanning electron and transmission electron microscopies. The results could be classified
into two distinct categories based on the laser energy density used. It was observed that the
thickness-dependent $E$ required to melt the Co film ($E_{Co}$) was lower than Si ($E_{Si}$)
primarily because of the higher reflectivity of Si. Consequently, for energy densities $E_{Co}<E_{1}<E_{Si}$
that preferentially melted the Co film, spatially ordered nanoparticles were formed and were
attributed to capillary-driven transport in the liquid phase. The ordering length scale corresponded
to the interference fringe spacing $\Lambda$ and the microstructure was primarily Co metal and
a metal-rich silicide phase. The native oxide layer played an important role in minimizing the
Co-Si reaction. For laser energy densities $E_{2}\geq E_{Si}$, spatially ordered patterns with
periodic length scales $L<\Lambda$ were observed and resulted from interference of an incident
laser beam with a beam transmitted into the Si substrate. These nanopatterns showed one- as well
as two-dimensional spatial ordering of the nanostructures. The microstructure in this laser energy
regime was dominated by silicide formation. These results suggest that various nanostructures
and microstructures, ranging from nearly pure metal to a Si-rich silicide phase can be formed
by appropriate choice of the laser energy simultaneous with film growth. 
\end{abstract}

\section{Introduction}

Metal and metallic compounds with nanometer scale (1-1000 nm) features arranged in spatially
ordered geometries offer various functionalities \citep{inomata98,stahl96}, including: in regular
arrays of discrete nm scale magnets for potential memory storage devices \citep{white00,black00};
surface plasmon waveguides made from linear chains of metal nanoclusters \citep{quinten98,maier03};
and as spatially ordered metal nanocatalyst seeds to make high-efficiency flat-panel displays
from carbon-nanotube arrays \citep{fan99}. In these applications, the control of ordering, including
spatial arrangement of the nanofeatures, control of its size, shape, composition, and crystallographic
orientation is important. In most instances, the ability to control ordering over large length
scales, i.e. to have long-range ordering, is also necessary \citep{lodder02}. In this regard,
laser-based patterning of surfaces of metals, semiconductors and insulators is a promising approach.
The area of laser-surface interactions has been broadly explored in the past for a fundamental
understanding of pattern forming processes and towards applications requiring ordered surface
structures \citep{bauerle96,miller98}. One area is that of laser-induced chemical vapor deposition
\citep{ehrlich83} while the other is direct irradiation of surfaces leading to rippling and
pattern formation due to various mechanisms, including interference between incident and scattered
light waves \citep{birnbaum65,Brueck82}, thermocapillary flow \citep{cline81} and other instabilities
\citep{Arnold95}. This pattern forming approach has been used in various applications, including
grating creation \citep{kim95} and quantum dot arrays \citep{kumagai93}. Recently, the group
of Pedraza et al. \citep{pedraza04a,pedraza04b} showed how spatially ordered nanostructures
could be formed in controlled ways by irradiation of the Si surface with ns laser pulses. The
patterning process was a result of deformation of the surface via thermocapillary effects, while
the length scales observed in the patterning also included those arising from interference of
the incident light with scattered waves. 

Another interesting area of pattern formation is when laser light interacts through optical and/or
thermal effects with a time-varying system, such as found in the deposition of thin films. During
the early stages of film growth from the vapor phase, the surface morphology changes dramatically
due to nucleation and growth processes. These processes are in turn critical functions of experimental
parameters like deposition rate, surface temperature and the presence of surface defects. In
this respect, a large body of work has been performed on developing strategies to create thin
film structures with homogeneous properties over large areas, for example having uniformity in
thickness, crystal structure, grain orientation, etc. in the plane of the substrate \citep{matthews75}.
On the other hand, recent interest in creating large-area structure having spatially ordered
inhomogeneities on the nm length scale have spurred new experiments to modify film nucleation
and growth. One example is the recent work of Timp and co-workers \citep{timp92,behringer97b}
and McClelland and co-workers \citep{mcclelland93} who used a laser-interference pattern to
modify atom beams such that the flux of depositing film material was spatially periodic on the
substrate surface. Consequently, nm scale periodic patterns were generated. 

Recently we investigated the role of thermal effects arising from laser-interference irradiation
performed simultaneously with metal film deposition. We showed in these experiments that when
the growing film/substrate surface is simultaneously irradiated with ns pulsed laser interference
patterns, long-range ordered morphologies can be achieved \citep{zhang03,zhang04}. In those
experiments, a 9 ns pulsed laser operating at 50 Hz with energy density $\leq20\, mJ/cm^{2}$
were used to dynamically pattern the films and the mechanism was attributed primarily to surface
diffusion of the metal atoms under the transient temperature gradients created on the substrate
surface. However, since the temperature transients only lasted for times of the order of 10 ns
\citep{Trice06} which was very small compared to the time between pulses of 20 ms, the observed
patterning was weak, as evidenced by the diffuse lines with small aspect ratio \citep{zhang04}.
Many nanoscale applications require well defined surface structures such as wires and particles
which have aspect ratios of $O(1)$. We recently observed that when dynamic patterning is performed
under laser energy densities that are sufficient to melt the metal film, then features with larger
aspect ratios can be achieved \citep{zhang05a}. 

In this work we present detailed experimental results of dynamic patterning of nanoscopic Co
on Si(001) as a function of film thickness and laser energy in regimes dominated by melting.
The patterning was performed using two-beam interference irradiation with a 9 ns, 266 nm pulsed
laser operating at 50 Hz. The resulting nanostructure and microstructure were characterized using
atomic force microscopy (AFM), secondary electron microscopy (SEM) and transmission electron
microscopy (TEM).  We determined that the laser melt threshold $E_{m}$ for the bare Si surface
was larger than for Co films, $E_{Co}$, primarily due to the difference in reflectivities. Also,
$E_{Co}$ increased as the film thickness decreased due to the enhanced heat sinking by the substrate.
As a result of this, the patterning results could be classified into two categories: (i) For
patterning in regime $E_{Co}<E_{1}<E_{Si}$, the patterns showed a length scale consistent with
the interference fringe spacing $\Lambda$ and the microstructure was primarily Co metal and/or
a Co-rich phase. The pattern mechanism in this regime was attributed primarily to capillary flow
in the liquid phase based on an estimate of the time scale and flux; (ii) The patterning in regime
$E_{2}\geq E_{Si}$ was dictated primarily by the rippling of the Si surface with the ordering
length scales having spacing $L<\Lambda$ because of interference between an incident beam and
a transmitted electromagnetic wave in Si. The microstructure in this regime was dominated by
silicide formation. These results show that a wide variety of nanostructures and microstructures
can be obtained by choice of laser energy and film thickness.

\section{Experiment\label{sec:Experiment} }

Co films ranging from 0.5 nm to 10 nm in thickness were deposited by e-beam evaporation onto
p-type Float-zone Si(001) substrates with resistivity of $65-80\,\Omega-cm$. E-beam evaporation
of Co was performed at room temperature in ultrahigh vacuum (base P of $2\times10^{-8}Torr$)
at deposition rates between 0.1 to 1 nm/min. A \emph{TECTRA} e-flux mini e-beam evaporator was
used for this purpose. The evaporation and laser irradiation geometry in shown in Fig. \ref{cap:figSchematic}.
\emph{In-situ} thickness measurements were made using an \emph{Inficon} Model XTM/2P quartz crystal
thickness monitor. Roughness measurements were made of the deposited films using the AFM. The
measurements showed that average peak-to-peak roughness was less that 0.2 nm for the range of
thickness studied and virtually independent of the deposition rate. The deposition chamber, laser
and necessary laser optics were mounted on a custom-built mechanical isolation table with I-2000
isolation legs from Newport. The vacuum pumping was achieved by a combination of an \emph{Osaka}
Magnetically levitated turbomolecular pump model TH260MCA and \emph{Physical Electronics} TiTan
400L Ion pump. The backing pump was a \emph{Leybold} ECODRY-L oil-free pump located in an adjacent
room to minimize vibration effects. The Si(001) substrates were degreased in methanol, acetone
and DI water prior to insertion into the chamber. We observed that the removal of the native
oxide layer enhanced formation of reacted microstructures and hence no attempt was made to etch
the oxide for the results presented here. 

For the laser interference we used a \emph{Spectra Physics} Injection seeded Lab-130-50 Nd:YAG
laser operating at its 4$^{\textrm{th}}$ harmonic of 266 nm, with a temporal FWHM of $\tau_{p}$
$\sim$ 9 ns, repetition rate f = 50 Hz and spatial coherence length > 2 m to perform two-beam
laser interference irradiation. The unfocused output of the laser has a multi-mode profile with
linear polarization and an area of $1\, cm^{2}$ with an energy distribution that corresponded
to a $\sim70\%$ Gaussian fit. The maximum total energy per pulse was approximately 44 mJ. For
this experiment, the pattern formation reported was for regions of the beam in which the energy
variation was $\leq15\%$. Two-beam laser interference was performed using two p-polarized beams
obtained by splitting the original beam using a 50-50\% beamsplitter. The energy of each beam
was measured using an \emph{Ophir} optronics 30A-P thermopile head and the \emph{Nova} power
meter. For the interference, beam 1 was incident on the substrate at 45$^{\text{o}}$ incidence
while beam 2 was rotated by an angle of between $38^{o}-45^{o}$ from beam 1, as depicted in
Fig. \ref{cap:figSchematic}. To maximize the interference intensity contrast, the individual
beam energy densities were controlled by lenses. The primary asymmetry in the energy of two beams
interfering on the surface came from the higher reflectivity and the larger projected area for
beam 2. This was partially compensated by focusing the beams using a 500 mm lens in the path
of the off-normal beam and a 750 mm lens for the $90^{o}$ beam. We estimated that the experimental
fringe contrast was between $80-90\%$. The energy density of  \emph{each} beam after focusing
was in the range of 50-200 $mJ/cm^{2}$ for energy regime $E_{1}$ and $200-250\, mJ/cm^{2}$
for regime $E_{2}$. The results presented in this work were from an approximately 25 $\mu m$
$\times$ 25 $\mu m$ wide region of interference on the samples. The interference angle $2\theta$
between the beams varied between $38^{o}\,-45^{o}$ for the pair of vacuum ports chosen in this
work. The resulting interference length scale on the surface was $\Lambda=\frac{\lambda}{2Sin2\theta}$,
where $\lambda=266\, nm$ is the incident wavelength. 

Following the deposition and irradiation, both done in vacuum, the samples were removed for characterization.
The nanostructures were characterized using contact-mode atomic force microscopy (AFM) and/or
Scanning electron microscopy (SEM). AFM was performed in ambient conditions using a \emph{Molecular
Imaging} PicoScan AFM with a BudgetSensors AFM tip of spring constant 0.2 N/m. The measured lateral
($x-y$) resolution was $\sim$ 10 nm and $z$-height resolution was $\sim$ 0.2 nm. SEM was
performed using a Hitachi S-4500 microscope. The microstructure was studied in the TEM using
plan-view geometry and a 200 keV JEOL 2000FX microscope. The plan-view TEM samples were prepared
using a well-known Si etching procedure. The Si substrate was mechanically ground until it was
reduced to 100 micron thickness. Next an etchant of hydrofluoric acid, nitric acid and acetic
acid in a 5:2:2 ratio was used to etch through the back of the Si substrate till a hole was visible
at which point the sample was sufficiently thin to allow for TEM analysis.

\section{Results and Discussion}

We observed that the overall results on the nature of spatial ordering, resulting surface morphology
and microstructure could be separated into three distinct categories based on the laser energy
used. One category belonged to the regime when the laser energy density was above the critical
laser energy density $E_{Si}$ of $\sim200\, mJ/cm^{2}$. For all $E\geq E_{Si}$ we observed
that the bare Si surface underwent rippling under otherwise identical conditions of interference.
It has been shown earlier by Pedraza and co-workers that this surface rippling occurs when the
energy absorbed is sufficient to melt the Si surface \citep{pedraza04a,pedraza04b}. Therefore
the energy $E_{Si}$ was taken as the melt threshold for the Si surface under the experimental
conditions used in this work. Another category of results corresponded to energies $50\leq E_{1}<200$
where Si rippling could not be observed and the resulting Co morphology consisted of well-defined
nanostructures with clear evidence of a droplet microstructure. The lower energy identified as
$E_{Co}$,  was the lower bound on the threshold of melting of Co during the dynamic patterning.
The final category of surface structures observed has been detailed by us in earlier publications
and consisted primarily of diffuse morphologies at energies $E<E_{Co}$ \citep{zhang03,zhang04}.
Below we detail the results obtained in the two energy regimes $E_{1}$ and $E_{2}$. First we
begin by showing that the melt threshold energy for Co on Si is always smaller than the Si melt
threshold consequent to which the mechanism of pattern formation in the various regimes may be
identified.

\subsection{Estimate of melt threshold vs. Co film thickness and lateral temperature gradient}

In the experimental results presented in Sec. \ref{sub:Morphology in E1} we observed two important
results: (i) the Co film appeared to melt at energies lower than the Si substrate despite the
higher melting point of Co; and (ii) ordered patterns with length scales corresponding to the
laser interference fringe spacing were formed. To understand these two effects we performed analytical
and finite element calculations of heating by laser irradiation. We have recently derived an
analytic and numerical solution for  heating of nanoscopic metal films by a spatially uniform
pulsed laser on transparent and absorbing substrates \citep{Trice06}. Excellent agreement between
the solution and experiment was found in predicting the threshold energy density needed to melt
nanoscopic Co films (with thickness less than the laser absorption depth $\alpha^{-1}\sim11\, nm$)
on SiO$_{\text{2}}$ substrates. One of the important results obtained by us was that the laser
energy required to melt the film increased as the film thickness decreased for metal films on
\emph{insulating} substrates \citep{matthias94}. Here, the experiments were performed primarily
on thermally conducting Si substrates. To determine the trend in energy vs film thickness, we
computed the laser energy required to achieve a temperature rise of the Co metal film to the
melting temperature of Si of 1683 K by a uniform laser intensity distribution. In this way complications
of phase change could be avoided. Fig. \ref{fig:Schematic-illustrating-the} shows the geometry
of the film-substrate system used in the thermal modeling. The previously determined solution
for the local film heating was as follows \citep{Trice06}:\begin{equation}
T(t)=T_{o}+A-B\label{eq:AbsorbSub}\end{equation}

where: \[
A=\frac{2}{\sqrt{\pi}}\cdot\left(\frac{S_{1}}{K}+\frac{S_{2}}{a_{s}\sqrt{\alpha_{s}}}\right)\sqrt{t}+\frac{S_{1}}{K^{2}}\cdot\left(e^{K^{2}t}\cdot erfc(K\sqrt{t})-1\right)\]

\[
B=\frac{S_{2}}{a_{s}^{2}\alpha_{s}K}\cdot\left(K+a_{s}\sqrt{\alpha_{s}}+\frac{K^{2}\cdot e^{a_{s}^{2}\alpha_{s}t}erfc(a_{s}\sqrt{\alpha_{s}t})}{a_{s}\sqrt{\alpha_{s}}-K}-\frac{a_{s}^{2}\alpha_{s}e^{K^{2}t}erfc(K\sqrt{t})}{a_{s}\sqrt{\alpha_{s}}-K}\right)\]

\noindent $S_{1}$ and $S_{2}$ are defined via\begin{equation}
S_{1}=\frac{(1-R(h))}{(\rho C_{p})_{m}h}\cdot\frac{E}{t_{p}}\cdot(1-exp(-a_{m}h))\label{eq:FilmSource}\end{equation}
\begin{equation}
S_{2}=\frac{(1-R(h))\cdot a_{s}}{(\rho C_{p})_{s}}\cdot\frac{E}{t_{p}}\label{eq:SubSource}\end{equation}

\noindent $\alpha_{s}$ is expressed as \begin{equation}
\alpha_{s}=\frac{k_{s}}{(\rho C_{p})_{s}}\label{eq:Subthermal}\end{equation}

\noindent and $K$ is given by\begin{equation}
K=\frac{\sqrt{(\rho C_{p}k)_{s}}}{(\rho C_{p})_{m}h}\label{eq:K}\end{equation}

\noindent Here $T_{o}$ corresponds to room temperature; R(h) is the effective film-thickness
dependant reflectivity for the Co/Si system as proposed by Ref. \citep{Heavens55}. A plot of
$R$ versus $h$ is presented in Fig. \ref{fig:Threshold}(a) for the Co-Si system; $E_{o}$
is the laser energy density; $a$ is the material absorption coefficient; $\rho$ is the density;
$C_{p}$ is the specific heat; $h$ is the Co film thickness; $t_{p}$ is the laser pulse width;
and $k$ is the thermal conductivity. The subscripts {}``m'' and {}``s'' denote the metal
and the substrate respectively. In Fig. \ref{fig:Threshold}(b), the predicted trend in the needed
threshold energy density to heat the Co film to the Si melting point is presented. Since our
experiments primarily involved two-beam interference, we have presented the figure in terms of
the energy density of an individual beam with the assumption of identical energy in the second
beam and therefore a resulting fringe contrast of 100\%. As the figure shows, the energy required
to melt a continuous Co film increases with decreasing film thickness. This trend arises because
for two primary reasons: (i) the increasing reflectivity of the Co/Si system as a function of
decreasing film thickness; and (ii) the increasing role of the Si substrate as a heat sink. Based
on this calculation, our experimental observations, to be discussed in sec. \ref{sub:Morphology in E1}
and \ref{sub:Morphology in E2}, appear to be consistent with patterning in regime $E_{1}$ arising
from preferential melting of Co and in $E_{2}$ due to the melting of Si respectively. The second
calculation made by us was an estimate of the lateral temperature gradient along the plane of
the film arising from the laser interference pattern. We used the finite element method via the
commercial code FlexPDE. The model solved the two-dimensional heat equation in the film and substrate
domains. The heat equation in the film domain was given by:\begin{equation}
\frac{\partial T(x,y,t)}{\partial t}=\frac{k_{m}}{(\rho C_{p})_{m}}\left(\frac{\partial^{2}}{\partial x^{2}}T(x,y,t)+\frac{\partial^{2}}{\partial y^{2}}T(x,y,t)\right)+4\cdot S_{1}^{\prime}cos(\frac{2\pi x}{F})\cdot\exp(-a_{m}\cdot y)\label{eq:NumFilm}\end{equation}

\noindent and in the substrate domain \begin{equation}
\frac{\partial T_{s}(x,y^{\prime},t)}{\partial t}=\frac{k_{s}}{(\rho C_{p})_{s}}\left(\frac{\partial^{2}}{\partial x^{2}}T_{s}(x,y^{\prime},t)+\frac{\partial^{2}}{\partial y^{2\prime}}T_{s}(x,y^{\prime},t)\right)+4\cdot S_{2}cos(\frac{2\pi x}{F})\cdot\exp(-a_{s}\cdot y^{\prime})\label{eq:Substrate_Num}\end{equation}

\begin{flushleft}where the origin of the $y^{\prime}$-coordinate system lies at the film-substrate
interface (Fig. \ref{fig:Schematic-illustrating-the}). Note that F is given by\begin{equation}
F=\frac{\lambda}{2\cdot sin(2\theta)}\label{eq:Fringespace}\end{equation}
\par\end{flushleft}

\noindent where $\lambda$ is the laser wavelength and $\theta$ is the interference angle in
radians. $S_{1}^{\prime}$ is given by \begin{equation}
S_{1}^{\prime}=\frac{(1-R(h))\cdot a_{m}}{(\rho C_{p})_{m}}\cdot\frac{E}{t_{p}}\cdot\label{eq:UnavergFilmHeatSource}\end{equation}
 The boundary condition for continuity of thermal flux at the interface is given by $k_{m}\frac{\partial T(x,y,t)}{\partial y}\mid_{y=h_{m}}=k_{s}\frac{\partial T_{s}(x^{\prime},y^{\prime},t)}{\partial y^{\prime}}\mid_{y^{\prime}=0}$.
Initially, the temperature of the system was at ambient and could be expressed by $T(x,y,t)\mid_{t=0}=T_{s}(x^{\prime},y^{\prime},t)\mid_{t=0}=T_{o}$.
To approximate infinite substrate thickness, we imposed the condition $T_{s}(x^{\prime},y^{\prime},t)\mid_{y^{\prime}=5\cdot L_{th}}=T_{o}$
where $L_{th}$ is the thermal diffusion length of the substrate given by $L_{th}=\sqrt{t_{p}\alpha_{s}}$.
To approximate infinite lateral extent, the period boundary conditions $T_{m}(x,y,t)=T_{m}(x+F,y,t)$
and $T_{s}(x^{\prime},y^{\prime},t)=T_{s}(x^{\prime}+F,y^{\prime},t)$ were also imposed. In
Fig. \ref{fig:Typical-results-of}, the calculated behavior of the temperature on the surface
of a continuous Co film on Si under laser-interference irradiation is shown. The Co film was
1 nm thick while the interference fringe spacing was 400 nm. In Fig. \ref{fig:Typical-results-of}(a)
the instantaneous temperature profile along the x-direction is shown at a time of 10 ns into
the irradiation and the temperature variation is periodic with the fringe spacing of $\Lambda=400\, nm$.
In Fig. \ref{fig:Typical-results-of}(b) the time dependence of the temperature at the fringe
maxima and minima is plotted. The phase changes, i.e. melting and resolidification of film and
substrate, are visible as distinct slope changes in the curves. From this behavior, the lateral
thermal profile as a function of time can be evaluated by subtracting the temperatures at the
positions of maxima and minima, as shown in Fig. \ref{fig:Typical-results-of}(c) From such numerical
calculations, the maximum lateral temperature difference was found to be in the range of $1-10\, K/nm$
$ $for Co films in the thickness range investigated in this work. 

\noindent

\subsection{Morphology, length scale and microstructure for energy $E_{Co}\leq E_{1}<E_{Si}$ \label{sub:Morphology in E1} }

Fig. \ref{fig:Spatial-orderingInterference} represents the typical results following patterning
in energy regime $E_{1}$. Fig. \ref{fig:Spatial-orderingInterference}(a) shows a SEM micrograph
and its power spectrum ($PS$) following dynamic patterning of a 4 nm Co film deposited at a
rate of 1 nm/min with each beam energy of $150\, mJ/cm^{2}$. The morphology consists of periodically
arranged rows of nanoparticles. The rows are regularly arranged with spacing $L=400\, nm$, which
was identical to the interference fringe spacing $\Lambda$. This regular arrangement had long-range
periodicity as confirmed by the computer generated diffraction spots visible in the $PS$. Also,
the direction of the pattern wavevector was identical to the fringe wavevector. The arrangement
of the nanoparticles within each row appeared random, as evidenced by the diffuse intensity in
the $PS$ for directions perpendicular to the diffraction spots. Fig. \ref{fig:Spatial-orderingInterference}(b)
shows a TEM micrograph and corresponding selected area diffraction (SAD) pattern of a dynamically
patterned 6 nm Co film deposited at a rate of 1 nm/min. The LRO period is consistent with the
fringe spacing of 400 nm. Indexing of the spots and rings of the diffraction pattern indicated
that the crystalline phase of the nanostructures could be indexed primarily as Co metal with
bulk hcp phase and the metal-rich silicide phase Co$_{\text{3}}$Si. Table \ref{tab:PhaseInterf}
shows the measured spots and rings corresponding to the nanostructures and they are in excellent
agreement with the theoretical spacings for bulk hexagonal Co and Co$_{\text{3}}$Si phases. 

Fig. \ref{fig:ThickDepInterf} shows the result of dynamic patterning with single beam energy
of $180\, mJ/cm^{2}$ and fringe spacing $\Lambda=400\, nm$ for films of thickness ranging from
0.5 nm to 25 nm . The important feature was that the LRO length scale and wavevector direction
corresponded to the interference fringe. However, another important result was that as the particle
sizes increased and became comparable to the line spacing $\Lambda$, an increasing interaction
between the lines was visible, as evident for the 15 nm and 25 nm films in Fig. \ref{fig:ThickDepInterf}(b)
and (c) respectively. This interaction led to a decrease in the overall quality of the spatial
order, as evident from the decreasing contrast of the diffraction spots in the $PS$ for the
various thickness. We also observed that the state of the Si surface played a critical role on
the observed pattern microstructure. Identical dynamic patterning experiments were performed
on Si surfaces treated by an HF-etch to remove the native oxide. Under these conditions we observed
that while the patterns still showed the interference fringe length scale, the morphology was
dominated by a reaction-like microstructure. Fig.  \ref{fig:Role-of-the-oxide} shows such a
reaction-dominated microstructure in contrast to the particle-like morphology of Fig. \ref{fig:Spatial-orderingInterference}.

\subsubsection{Mechanism of pattern formation in regime $E_{1}$ \label{sub:Mechanism-of-pattern}}

We attribute the pattern formation in regime $E_{1}$ to be dominated by mass transport in the
liquid phase with  contribution from solid state mass transport during the early stages of deposition.
This is based on two experimentally observed features: (i) The final observed nanomorphology
shows particulate formation only for energies in regime $E_{1}$. We suggest that this is because
at some stage of the dynamic patterning the metal film underwent melting; and (ii) As observed
by us previously the patterns formed for energies $<E_{1}$ consist of diffuse lines with small
aspect ratio \citep{zhang03,zhang04}. Considering that the melt threshold of the Co metal film
is highly thickness dependent and increases with decreasing film thickness, as shown in Fig.
\ref{fig:Threshold}(a), we suggest that during the early stages of deposition and irradiation
the film does not melt. Rather, mass transport occurs via surface diffusion and surface self-diffusion
of Co leading to the nucleation and growth of metal clusters. Fig. \ref{fig:Model-for-patterning}
depicts this model, where after early stages of growth the film consists of discrete particles
with larger average size in the lower temperature regions of the fringe. Subsequently when the
clusters reach a thickness wherein the laser energy can melt it, liquid phase mass transport
will be important. This sequence of solid-state transport and liquid transport eventually results
in the observed particulate like patterns. 

Based on the observations and the above hypothesis, we have identified the possible hydrodynamic
flow mechanisms leading to pattern formation based on the the time scales of  various contributions
to liquid motion. One is the non-uniform temperature of the metal liquid arising from the interference
pattern that will result in a surface tension gradient hence contributing a thermocapillary component
to flow. A second contribution is the Laplace pressure difference arising from the difference
in average metal cluster size that determines the morphology during early stages of film growth.
We further assume that the clusters are sufficiently large (i.e of few 10's of nm) such that
their melting and liquid flow contributing to pattern formation can be analyzed using a continuum
level approach. We can estimate the time scales involved by evaluating the Navier-Stokes equation
for incompressible 1-D flow of a liquid flowing in the $x$-direction given by:\begin{equation}
\rho\frac{dv_{x}}{dt}=-\frac{dp}{dx}+\eta\frac{d^{2}v_{x}}{dy^{2}}\label{eq:NS1}\end{equation}

where $\rho$ is the density of the liquid; $v_{x}$ is the $x$-component of the liquid velocity;
$\frac{dp}{dx}$ is the pressure gradient in the direction of flow $x$; and $\eta$ is the dynamic
viscosity. Solving for the low Reynolds number case, which is valid for such thin films, the
inertial contribution from $\rho\frac{dv_{x}}{dt}$ can be neglected (analogous to the steady
state condition) and we get on integrating: \begin{equation}
v_{x}=\frac{1}{2\eta}\frac{dp}{dx}y^{2}+ay+b\label{eq:velocity}\end{equation}

where a and b are constants of integration. Applying the no-slip boundary condition of $v_{x}=0$
at $y=0$, i.e. the liquid-substrate interface, we get $b=0$. The free surface boundary condition
at the liquid-vacuum interface requires that the tangential stress gradients cancel out and hence
$\sigma_{xy}=0$ at $y=h$, the local liquid height. In the interference experiments, the non-uniform
heating results in a non-uniform liquid temperature along the plane of the liquid surface $x$
 causing the surface tension to vary with $x$. Further, no temperature gradients exist in the
vertical $y$-direction because of the small film thickness and the large thermal conductivity.
Hence the stress balance condition at the surface, which requires that the viscous stress be
balanced by the surface tension gradient stress gives: \begin{equation}
\eta\frac{dv_{x}}{dy}=\gamma_{T}\frac{dT}{dx}\, at\, y=h\label{eq:Stress-balance}\end{equation}
where the $LHS$ is the viscous stress; $\gamma_{T}$ is rate of change of surface tension with
temperature; and $\frac{dT}{dx}$ is the gradient in liquid surface temperature in the x-direction.
Using eq. \ref{eq:Stress-balance} in eq. \ref{eq:velocity} we get: \begin{equation}
v_{x}=\frac{1}{2\eta}\frac{dp}{dx}y^{2}-\frac{1}{\eta}\frac{dp}{dx}hy+\frac{\gamma_{T}}{\eta}\frac{dT}{dx}y\label{eq:v_x2}\end{equation}

The effective liquid velocity can be obtained by averaging over the local thickness of the film
such that \citep{kondic03}:\begin{equation}
v_{av}=\frac{1}{R}\int_{o}^{R}v_{x}dy=-\frac{R^{2}}{3\eta}\frac{dP}{dx}+\frac{R\gamma_{T}}{2\eta}\frac{dT}{dx}\label{eq:v_av}\end{equation}

To estimate the time scale of the liquid flow, the form of the pressure gradient must be explicitly
introduced. In our experiments, the dynamic patterning process involves the constant deposition
of metal with irradiation by the $9\, ns$ pulses spaced $20\, ms$ apart. Numerical calculations
show that the liquid lifetime ranges from 1 to 10 ns and therefore, each pulse is thermally independent
\citep{Trice06}. In addition, the laser-metal interaction time is a small fraction of the deposition
time. Therefore, we can simplify the analysis by analyzing the two likely contributions to pattern
formation. (i) During the early stages of deposition when the atom clusters are extremely small
and the film is discrete, the non-uniform laser heating produces a small surface height variation
$h(x)$ by surface diffusion, as observed by us in previous work \citep{zhang03,zhang04}. The
height variation is manifested as a film morphology that consists of metal nanoclusters of various
sizes R(x), with the larger clusters in the thicker regions (Fig. \ref{fig:Model-for-patterning}).
In addition, the height variation has the periodicity of the laser fringe $\Lambda$. We further
assume that the surface-diffusion results in larger clusters in the regions of the interference
\emph{minima}, as would be expected for a thermal-gradient assisted diffusion of surface atoms.
(ii) The interference-melting of this non-uniform height leads to the eventual observed nanoparticle
patterns. To show this, we use the fact that the metal nanoclusters of various radius of curvature
$R(x)$ producing the height variation will contribute a local Laplace pressure given by $-\gamma(x)K$,
where $K=\frac{d^{2}h}{dx^{2}}=-\frac{1}{R(x)}$ is the local curvature, which is positive for
spherical shaped particles. The surface tension is also now a function of film position because
of the thermal gradients resulting from the fringe. Therefore the pressure gradient in the $x$-direction
can be expressed as:  \begin{equation}
\frac{dp}{dx}=-\gamma\frac{dK}{dx}-K\frac{d\gamma}{dx}=-\frac{\gamma(x)}{R^{2}}\frac{dR}{dx}+\frac{\gamma_{T}}{R(x)}\frac{dT}{dx}\label{eq:GradP}\end{equation}

where we have approximated the local curvature to be given by $K=1/R(x)$. The resulting flow
velocity at any position x can be obtained by using the above equation in eq. \ref{eq:v_av}
giving: \begin{equation}
v_{av}=\frac{\gamma}{3\eta}\frac{dR}{dx}-\frac{1}{6\eta}\gamma_{T}\frac{dT}{dx}R\label{eq:v_x3}\end{equation}
where we have approximated the local film height $h\sim R(x)$. The first term contributes to
the flow a component similar to that found in Ostwald ripening where regions with lower Laplace
pressure grow at the expense of regions with higher pressure. The second term $\gamma_{T}\frac{dT}{dx}$
is always negative because the surface tension decreases with increasing pressure and therefore
contributes a thermocapillary enhancement to the Laplace component of the pressure (1$^{\text{st}}$
term in Eq. \ref{eq:v_x3}). Equation \ref{eq:v_x3} reduces to the classical Marangoni or thermocapillary
flow for a flat film when $\frac{dR}{dx}=0$ and $R=h$. The time scale $\tau_{L}$ of liquid
motion over the characteristic length scale of the experiment, given by the distance between
the fringe maxima to minima (length scale of $\Lambda/2$) can be estimated by noting that $v_{av}=$$\frac{\Lambda/2}{\tau_{L}}$
and so:\begin{equation}
\tau_{L}=\frac{\Lambda}{2}\left\{ \frac{1}{\frac{1}{6\eta}\mid\gamma_{T}\frac{dT}{dx}\mid R+\frac{\gamma}{3\eta}\frac{dR}{dx}}\right\} \label{eq:Tau_L}\end{equation}

To make an order of magnitude estimate of $\tau_{L}$, we used the following numbers: $\eta^{Co}=4.46\times10^{-3}Pa-s$;
$\gamma^{Co}=1.88\, J/m^{2}$; $\mid\gamma_{T}^{Co}\mid=0.5\times10^{-3}J/m^{2}-K$; and the
maximum transient thermal gradient in the fringe as estimated from numerical simulations of $dT/dx\sim10\, K/nm$.
 An estimate of $R$ and the thickness gradient is more difficult. However, we make two approximations:
for the value of $R$ we used the average final deposited film height of $h_{f}$ while for $dR/dx$
 we used our previous results \citep{zhang03,zhang04} which showed that the Co lines resulting
from surface diffusion lead to a height approximately 15-20 times that of the final average film
thickness $h_{f}$ resulting in  $dR/dx\sim20\frac{h_{f}}{\Lambda/2}$. We find then that the
above equation yields two important results: (i) the Marangoni term is small compared to the
capillary term. This suggests that the Laplace pressure difference due to difference particle
size plays a dominant role and the patterning process is similar to Ostwald ripening, i.e. the
larger particles grow at the expense of the smaller ones; and (ii) the expression can be simplified
to $\tau_{L}=\frac{3\eta\Lambda^{2}}{80\gamma h_{f}}\sim8.89\times10^{-5}\frac{\Lambda^{2}}{h_{f}}$
which varies inversely as the film thickness. We can then estimate the time scale for liquid
motion occurring at any stage when the surface diffusion has established a thickness variation
and when the average film thickness is sufficient such that the film melts under the laser pulse
process For instance, if the film melts at a thickness of $h_{f}\sim0.5\, nm$ we get a time
scale of $\sim28\, ns$ for a fringe spacing $\Lambda=400\, nm$. While this time implies that
the patterning cannot not occur under one pulse, it is achievable due to the long integrated
liquid lifetime over the time scale of the experiment. The typical experiment times ranged from
0.5 minutes to 5 minutes at a repetition rate of 50 Hz. Using an average  liquid lifetime of
$\sim5\, ns$ \citep{Trice06}, the total integrated liquid lifetime can be considerable because
between 1500 to 15000 pulses are incident during the deposition process.. Therefore, even taking
into account that during the initial pulses, when mass transport via solid state processes are
likely to dominate, a sufficient integrated time is available for liquid phase motion to dominate
the pattern formation.

One can estimate whether mass transport via surface diffusion in the solid state can make a substantial
contribution to pattern formation throughout the entire experiment. An estimate of the time scale
as well as the efficiency of mass transport for surface diffusion can be made using the theory
given by Mullins \citep{mullins57} in which the time scale and flux for mass transport are respectively:\begin{equation}
\tau_{S}=\frac{\Lambda R^{2}}{2\frac{D_{s}}{kT}\gamma\Omega\frac{dR}{dx}}\label{eq:tau_S}\end{equation}
\begin{equation}
J_{s}=-\frac{2D_{S}\gamma}{kTR^{2}\lambda}\Omega\nu\frac{dR}{dx}\label{eq:J_S}\end{equation}

where, $D_{s}$ is the surface self-diffusion coefficient; $\Omega$ is the atomic volume; $k$
is Boltzmann's constant; and $\nu$ is the areal density of surface atoms. From this we can claim
that mass transport by the hydrodynamic process will dominate if: \begin{equation}
\frac{\tau_{L}}{\tau_{s}}<1;\,\, and\,\,\frac{J_{L}}{J_{s}}>1\label{eq:LiqvsSurf}\end{equation}

Using eq. \ref{eq:Tau_L} (and neglecting the Marangoni component) and eq. \ref{eq:tau_S} we
get at the melting point of Co of $T=1768\, K$: \begin{equation}
\frac{\tau_{L}}{\tau_{S}}=\frac{2D_{s}\Omega\eta}{kTR^{2}}=4\times10^{-12}\frac{D_{S}}{R^{2}}\label{eq:Tau_ratio}\end{equation}
 where we have used $\Omega=1/\rho_{at}^{Co}=1.1\times10^{-29}m^{3}/atom$, where $\rho_{at}^{Co}=8.97\times10^{28}\, atoms/m^{3}$
is the atomic density of Co. Therefore we see that for the liquid process to be faster  $\frac{D_{s}}{R^{2}}<2.5\times10^{11}m^{2}/s$.
Assuming a film thickness $h_{f}$ = R of 0.5 nm and using the experimentally observed value
for Co diffusivity of $D_{S}^{expt}=0.48\times10^{-8}e^{\frac{-0.14\, ev}{kT}}$ \citep{prasad84}
we see that $\frac{\tau_{L}}{\tau_{s}}\sim5\times10^{-8}$ and therefore the liquid time scale
is much shorter. However, the mass transport efficiency, given by the flux, must also be compared.
For the liquid flow, the flux in the $x$-direction can be expressed as: \begin{equation}
J_{L}=-v_{av}\rho_{L}^{at}\label{eq:J_liq}\end{equation}

where $\rho_{L}^{at}$ is the atomic number density in the liquid phase. From eq. \ref{eq:v_av},
eq. \ref{eq:J_S} and eq. \ref{eq:J_liq} we get that liquid transport dominates when:\[
\frac{J_{L}}{J_{S}}=\frac{(\rho_{Co}^{at})^{4/3}kTR^{2}}{6\eta D_{s}}\sim\frac{2\times10^{14}R^{2}}{D_{s}}>1\]
 where we have used the approximation that the liquid and solid atomic densities of Co are comparable,
i.e. $\rho_{L}^{at}\sim\rho_{Co}^{at}$. Here also we we find that for films of 0.1 nm thickness
or larger the flux carried by the liquid is significantly larger than the solid state implying
that liquid-based mass transport is most likely responsible for the observed pattern formation.

\subsection{Morphology, length scale and microstructure for energy $E_{2}\geq E_{Si}$ \label{sub:Morphology in E2}}

Fig. \ref{fig:Spatial-orderingSEW} represents the typical results following patterning in energy
regime $E_{2}$. Fig. \ref{fig:Spatial-orderingSEW}(a) shows an AFM topographic image and its
power spectrum ($PS$) following dynamic patterning of a 2 nm Co film deposited at a rate of
1 nm/min with an interference angle of $42^{o}$. The direction of the laser interference fringe
is marked in the top right corner of the AFM image. A distinctly different nanomorphology is
observed for this regime, as compared to Fig. \ref{fig:Spatial-orderingInterference}(a). The
quantitative differences include: (i) a more elliptical shape of the nanofeatures as compared
to the more droplet-like shapes observed in $E_{1}$; (ii) an ordering length scale different
from $\Lambda$; and (iii) a two-dimensionally (2-D) periodic arrangement of the features. The
LRO length scale in the direction of the interference wavevector, also seen as diffraction spots
in direction \emph{A} in the $PS$, was estimated to be 240 nm, which is smaller than $\Lambda=400\, nm$.
In addition, the features have spatial order along each of the rows, as evident from the 2-D
nature of the diffraction pattern in the $PS$. The spots visible in direction \emph{B} correspond
to a feature spacing of $\sim522\, nm$. Fig. \ref{fig:Spatial-orderingSEW}(b) shows a TEM micrograph
and corresponding selected area diffraction (SAD) pattern of a dynamically patterned 2 nm Co
film deposited at a rate of 1 nm/min. The general results, i.e. the 2-D form of ordering and
an LRO length scale of $L\neq\Lambda$ are consistent with the AFM image. Indexing of the spots
in the diffraction pattern indicated that the crystalline phase of the nanostructures was primarily
Cobalt silicide with excellent matching to the Si-rich CoSi$_{\text{2}}$ phase. Table \ref{tab:PhaseSEW}
shows the measured spot spacings and the theoretical spacings for cubic CoSi$_{\text{2}}$. 

Fig. \ref{fig:ThickDepInterf} shows the result of dynamic patterning in energy regime $E_{2}$
with fringe spacing $\Lambda=400\, nm$ for films of thickness of 0.5 nm (deposited at 0.1 nm/min)
and 1 nm (deposited at 1 nm/min). The 2-D ordering, as in the 2 nm case of Fig. \ref{fig:Spatial-orderingSEW},
is clearly evident from the AFM images and the corresponding $PS$. Again, like the 2 nm case
the LRO visible as spots in direction A, and had spacing $L<\Lambda$ with values of 232 nm and
233 nm for the 0.5 and 1 nm film respectively. Another feature is that the average LRO spacing
in the B direction increased with increased thickness and was estimated to be 172, 474 and 522
nm for the 0.5, 1 and 2 nm film. The average diameter of the features also increased and was
68, 98 and 120 nm for 0.5, 1 and 2 nm films respectively.

\subsubsection{Mechanism of pattern formation in regime $E_{2}$}

One of the striking results for irradiation in this regime is that pattern formation also occurred
on the bare Si substrate surface without any Co deposition. As shown in an exhaustive work by
Pedraza et al \citep{pedraza04a}, the rippling of the Si surface is due to melting and the various
resulting periodic length scales come from interference of beams scattered along the surface
due to surface roughness as well as with surface electromagnetic waves (SEW), a phenomenon that
has been widely studied in the past. Further, they also observed that the morphology was a function
of the irradiation time and proposed that hydrodynamic instabilities can cause the morphologies.
We suggest that our patterns observed in this energy regime come from similar mechanisms. The
first evidence is the origin of the observed LRO length scale in the \emph{A} direction of approximately
$\sim232\, nm$. The general grating wavelength $\Lambda^{g}$ resulting from interference between
the incident beam with wavevector $k$ and the scattered beam with wavevector $k^{sc}$ can be
expressed as: \begin{equation}
\Lambda_{+/-}^{g}=\frac{2\pi}{k^{sc}\pm kSin\theta_{i}}\label{eq:Lambda_Grating}\end{equation}

where $\theta_{i}$ is the incident angle of the laser light. Depending upon the mechanism of
the scattered beam, various grating wavelengths can be observed. In our case, the closest agreement
occurs for interference with the beam scattered in reflection for which case $k^{sc}=kn$, where
$n$ is the refractive index for Si at $266\, nm$ and is 1.831. Using this we find that $\Lambda_{-}^{g}=229\, nm$,
which is very close to the observed LRO visible as diffraction spots in the \emph{A} directions
of the PS of figures \ref{fig:Spatial-orderingSEW} and \ref{fig:ThickDepSEW}. In addition,
we observed an identical length scale on the bare Si surface suggesting that the deposition of
Co during the dynamic patterning process played a negligible role in determining the grating
spacing. This result is intriguing primarily because the patterned microstructure comprises a
Co-silicide phase while the LRO length scale is determined by the Si surface. The second ordering
length scale appearing in the \emph{PS} along direction \emph{B} (Fig. \ref{fig:Spatial-orderingSEW}
and \ref{fig:ThickDepSEW})\emph{,} which was perpendicular to the interference fringe direction,
cannot be explained by similar surface interference phenomenon because the length scale changes
with thickness and/or irradiation time. While further investigations are required to understand
this length scale in detail, one possible mechanism is a Rayleigh-like breakup of the elongated
liquid region consisting of the Co-Si mixture during the laser melting. It is well known that
the classical break-up of a liquid cylinder, driven by minimization of the surface area, leads
to preferred wavelengths and typically the wavelength and diameter increase linearly with the
cylinder diameter. Qualitatively, one would expect that as the Co concentration is increased
the volume of the Co-Si liquid phase will also increase and hence the observed trend of increase
in spacing and diameter of the particles in going from the 0.5 to 2 nm film.

\section{Conclusion}

We have investigated the pattern formation resulting from Co film deposition with simultaneous
ns pulsed laser interference irradiation of the Si(001) surface as a function of laser energy
and film thickness. We found that the $E$ required to melt the Co film ($E_{Co}$) was lower
than the bare Si surface ($E_{Si}$) because of the higher reflectivity of the Si surface and
the heat sinking effects of the substrate. Consequently, for energy densities $E_{Co}<E_{1}<E_{Si}$
that preferentially melted the Co film, spatially ordered nanoparticles were formed by capillary-driven
transport in the liquid phase. The ordering length scale corresponded to the interference fringe
spacing $\Lambda$ and the microstructure was primarily Co metal and a metal-rich silicide phase.
The native oxide layer played an important role in minimizing the Co-Si reaction. For laser energy
densities $E_{2}\geq E_{Si}$, spatially ordered patterns with periodic length scales $L<\Lambda$
were observed and resulted from interference of an incident laser beam with a beam transmitted
into the Si substrate. These nanopatterns showed both one- as well as two-dimensional spatial
ordering of the nanostructures. The microstructure in this laser energy regime was dominated
by Si-rich silicide phase formation. These results suggest that various nanostructures and microstructures,
ranging from nearly pure metal to the silicide can be formed by appropriate choice of the laser
energy simultaneous with film growth. 

RK acknowledges support by the National Science Foundation through a CAREER grant \# DMI-0449258
and is indebted to Prof. R. Sureshkumar for valuable comments on hydrodynamics. The authors would
also like to thank Chris Favazza and H.Krishna for confirming some of the experimental results.

\pagebreak

\section*{Figure captions:}

\begin{enumerate}
\item Schematic of the experimental approach used for \emph{dynamic patterning.} The deposition is
performed by electron beam evaporation while the two-beam laser interference irradiation is performed
using two p-polarized beams from a Nd:YAG laser operating at 266 nm wavelength and 9 ns pulse
width.\label{cap:figSchematic}
\item Schematic illustrating the film-substrate geometry used in the analytical and numerical calculations.
The boundary conditions at the various interfaces are indicated for the one beam and 2-beam interference
calculations. \label{fig:Schematic-illustrating-the}
\item (a) Plot of the thickness dependent reflectivity of the Co-Si based on the model of Heavens \citep{Heavens55}.
(b) Plot of laser energy density needed to bring Co-Si system to Si melting temperature expressed
in terms of the energy density of a single laser beam and assuming 100\% contrast between interfering
beams.   \label{fig:Threshold}
\item Typical results of numerical calculations of the laser interference heating of a continuous 1
nm Co film on Si substrate. (a) The temperature profile along the surface x-direction at 10 ns
showing the temperature variation is periodic with the fringe spacing of $\Lambda=400\, nm$.
(b) The time dependence of the temperature at the fringe maxima and minima. (c) The lateral thermal
profile as a function of time obtained from (b) by subtracting the temperature at any given time.
\label{fig:Typical-results-of}
\item Spatial ordering and microstructure following dynamic patterning at energy $E_{1}<E_{Si}$. (a)
SEM image and power spectrum of the pattern in a 4 nm Co film deposited at 1 nm/min. The nanoparticles
are arranged in rows which have long range order with a periodic length scale of 400 nm, which
is consistent with the interference length scale $\Lambda$. The LRO is visible as diffraction
spots in the power spectrum. No ordering is visible for the particles within each row, as evidenced
by the diffuse form of the power spectrum in directions perpendicular to the spots\emph{.} (b)
TEM micrograph and corresponding selected area diffraction pattern of a 6 nm Co film deposited
at 1 nm/min. The LRO period is consistent with the fringe spacing of 400 nm. Indexing of the
spots and rings, indicated by numbers on the diffraction image, suggest the presence of Co metal
and a small fraction of the metal-rich silicide phase Co$_{\text{3}}$Si (Table \ref{tab:PhaseInterf}).
\label{fig:Spatial-orderingInterference}
\item Pattern morphology imaged in the SEM and corresponding $PS$ following dynamic patterning in
energy regime $E_{1}$ as a function of film thickness. (a) 0.5 nm film deposited at 0.1 nm/min;
(b) 15 nm film deposited at a rate of 1 nm/min; (c) 25 nm film deposited at a rate of 1 nm/min.
In all three cases, the LRO length scale (400 nm) and wavevector direction corresponded to the
interference fringes. The average particle size increased with increasing film thickness and
therefore the nature of the ordering decreased presumably due to coalescence of the large particles
across the lines. This decrease in ordering quality is visible from the poorly defined diffraction
spots for the 25 nm film. Figures (a) and (b) are from our reference \citep{zhang05a} \label{fig:ThickDepInterf}
\item Role of the native oxide layer. SEM micrograph of patterning done in regime $E_{1}$ for Co on
Si surface following removal of the native oxide. The morphology shows evidence for a strong
reaction between the metal and substrate and is in striking contrast to the result of similar
patterning performed on the Si surface with the native oxide layer, as shown in Fig. \ref{fig:Spatial-orderingInterference}(a).
 \label{fig:Role-of-the-oxide} 
\item Model for the dynamic patterning process. During early stages of deposition and simultaneous
irradiation surface diffusion results in a film morphology described by metal clusters of varying
size, with the smaller clusters in the higher temperature regions of the laser interference fringe.
The clusters and the overlapping temperature profile due to the interference pattern are shown
in the figure. The final pattern results following melting and liquid motion from the smaller
clusters to the larger ones. \label{fig:Model-for-patterning}
\item Spatial ordering and microstructure following dynamic patterning at energy $E_{2}>E_{Si}$. (a)
AFM image and power spectrum of the surface nanostructure for a 2 nm Co film deposited at 1 nm/min.
The direction of the interference fringe is marked in the top right corner of the AFM image.
The surface features with periodicity in the direction consistent with the interference fringe
have a periodic spacing length scale of 270 nm, which is not consistent with the interference
length scale $\Lambda=400\, nm$. This period is visible as diffraction spots along direction
\emph{A} in the \emph{PS}. In addition, clear LRO is visible for the particles within each row
leading to a 2-dimensional periodic structure. This is also evidenced in the power spectrum,
with the additional diffraction spots visible along direction \emph{B}. (b) TEM micrograph and
corresponding selected area diffraction pattern of the dynamically patterned 2 nm Co film. The
2-D nature of the pattern is clearly visible. Indexing of the visible spots (Table \ref{tab:PhaseSEW})
indicates that the most dominant phase is the Si-rich silicide CoSi$_{\text{2}}$. \label{fig:Spatial-orderingSEW}
\item Pattern morphology imaged in the AFM with corresponding $PS$ following dynamic patterning in
energy regime $E_{2}$ as a function of film thickness. The line in the inset of the AFM figures
point along the expected 2-beam fringe direction. (a) 0.5 nm film deposited at 0.1 nm/min; (b)
1 nm film deposited at a rate of 1 nm/min. In both cases, a 2-D pattern is visible. Similar to
the 2 nm case (Fig. \ref{fig:Spatial-orderingSEW}) the LRO is visible as spots in the \emph{PS}
along direction \emph{A} and has spacing $L<\Lambda$ of 260 nm and 238 nm for the 0.5 and 1
nm film respectively. This length scale is similar to the 2 nm film of 249 nm. On the other hand,
the average feature size increased with increasing film thickness and the LRO indicated by spots
along \emph{B} was determined to be 172 and 474 nm respectively, as compared to 522 nm for the
2 nm film. \label{fig:ThickDepSEW}
\end{enumerate}
\pagebreak

\section*{Table Captions:}

\begin{enumerate}
\item Indexing of diffraction spots and rings for the SAD pattern in Fig \ref{fig:Spatial-orderingInterference}(b)
obtained in regime $E_{1}$. The best matching was obtained for crystalline phases corresponding
to the bulk Co hcp phase and the metal rich hexagonal Co$_{\text{3}}$Si silicide phase. \label{tab:PhaseInterf}
\item Indexing of diffraction spots and rings for the SAD pattern in Fig \ref{fig:Spatial-orderingSEW}(b)
obtained in regime $E_{2}$. The best matching was obtained for crystalline phases corresponding
to the Si-rich cubic CoSi$_{\text{2}}$ silicide phase. \label{tab:PhaseSEW}
\end{enumerate}
\pagebreak

\begin{figure}[!tph]
\begin{centering}\includegraphics[width=3in,keepaspectratio]{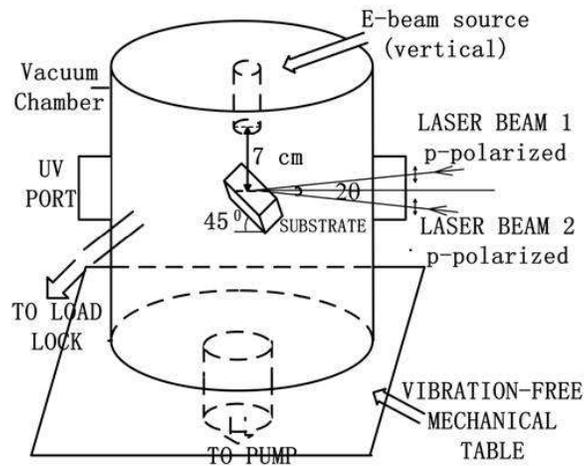}\par\end{centering}

\caption{Schematic of the experimental approach used for \emph{dynamic patterning.} The deposition is
performed by electron beam evaporation while the two-beam laser interference irradiation is performed
using two p-polarized beams from a Nd:YAG laser operating at 266 nm wavelength and 9 ns pulse
width.\label{cap:figSchematic}}
\end{figure}

\pagebreak

\begin{figure}[!tph]
\begin{centering}\includegraphics[height=3in,keepaspectratio]{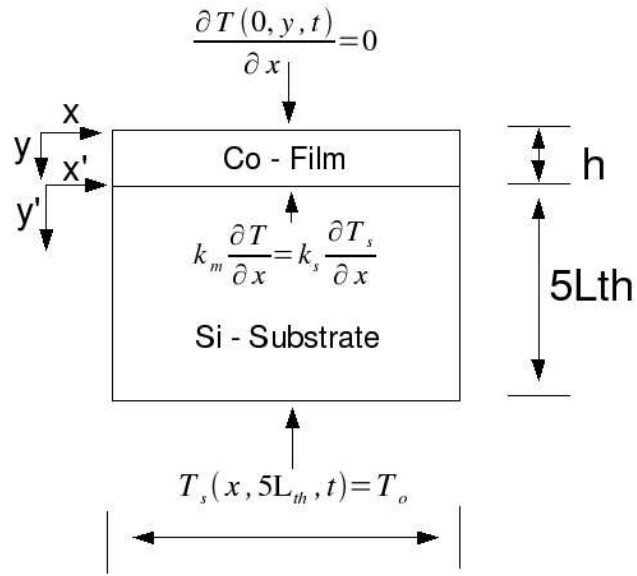}\par\end{centering}

\caption{Schematic illustrating the film-substrate geometry used in the analytical and numerical calculations.
The boundary conditions at the various interfaces are indicated. \label{fig:Schematic-illustrating-the}}
\end{figure}

\pagebreak

\begin{figure}[!tph]
\begin{centering}\subfigure[]{\includegraphics[width=2in,keepaspectratio]{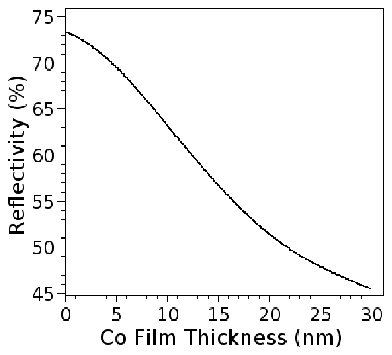}}\subfigure[]{\includegraphics[width=2in,keepaspectratio]{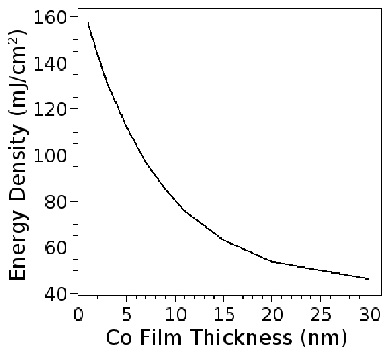}}\par\end{centering}

\caption{(a) Plot of the thickness dependent reflectivity of the Co-Si based on the model of Heavens
\citep{Heavens55}. (b) Plot of laser energy density needed to bring Co-Si system to Si melting
temperature expressed in terms of the energy density of a single laser beam and assuming 100\%
contrast between interfering beams.   \label{fig:Threshold}}
\end{figure}

\pagebreak

\begin{figure}[!tph]
\begin{centering}\includegraphics[width=2in,keepaspectratio]{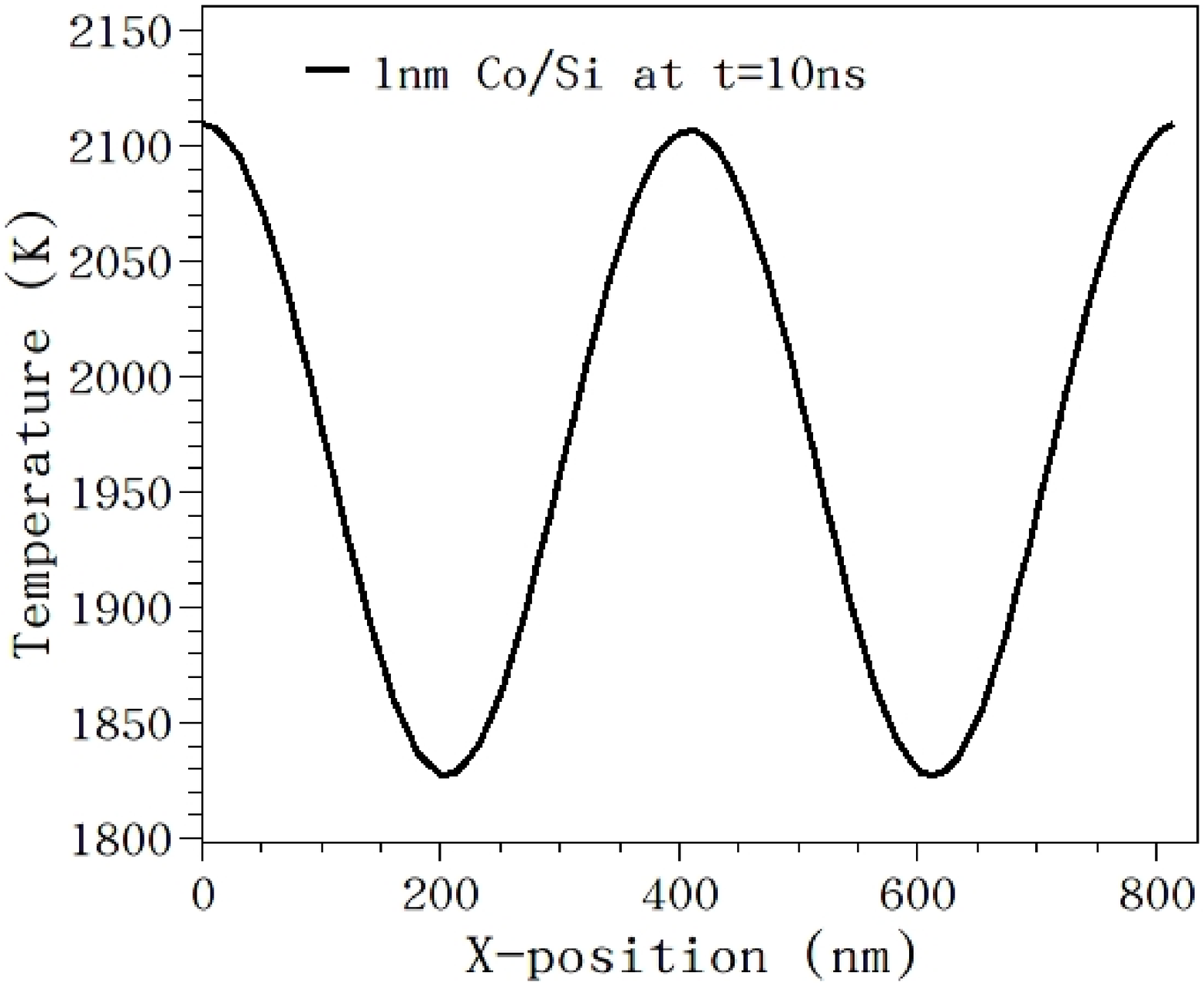}\subfigure[]{\includegraphics[width=2in,keepaspectratio]{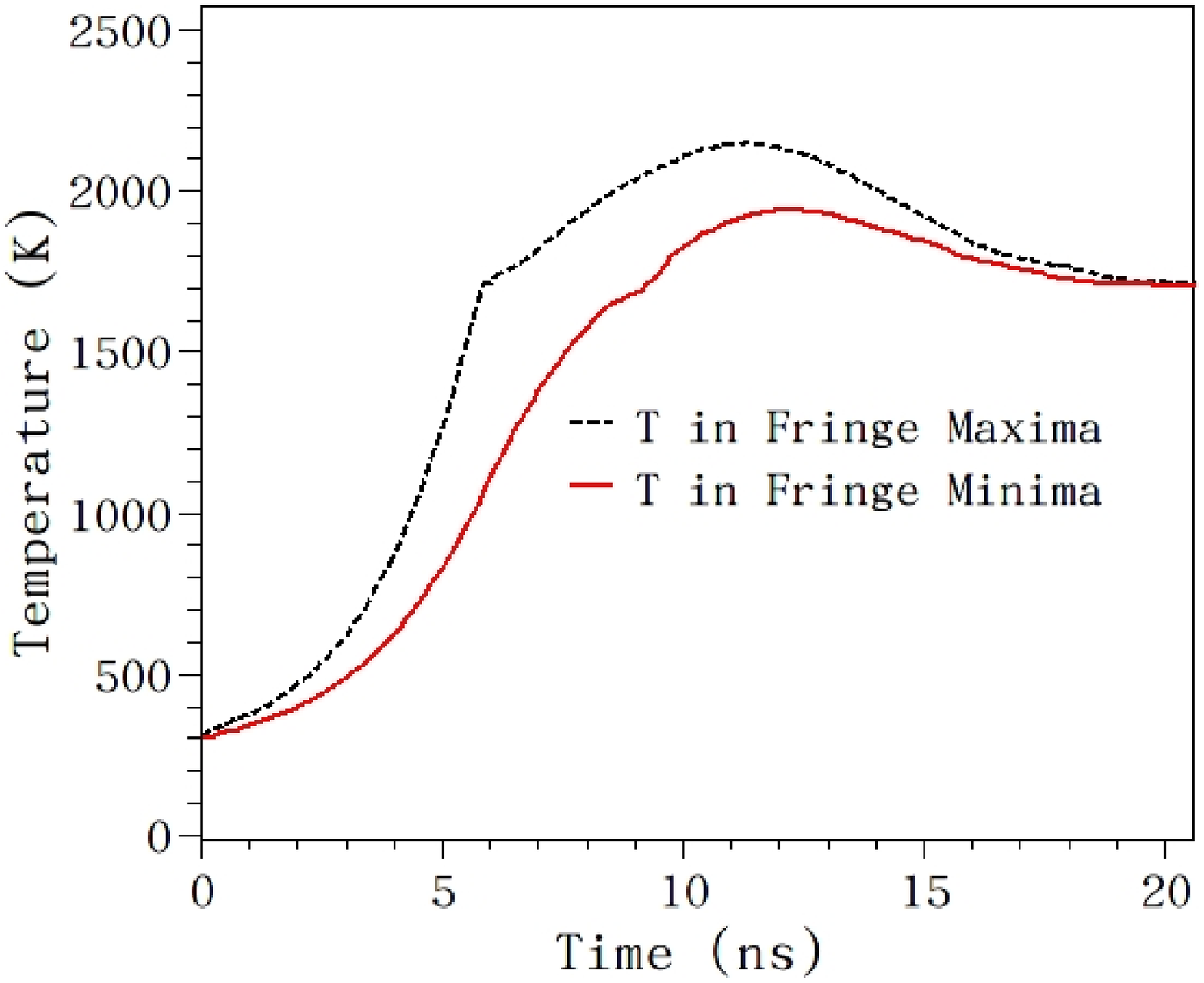}}\includegraphics[width=2in,keepaspectratio]{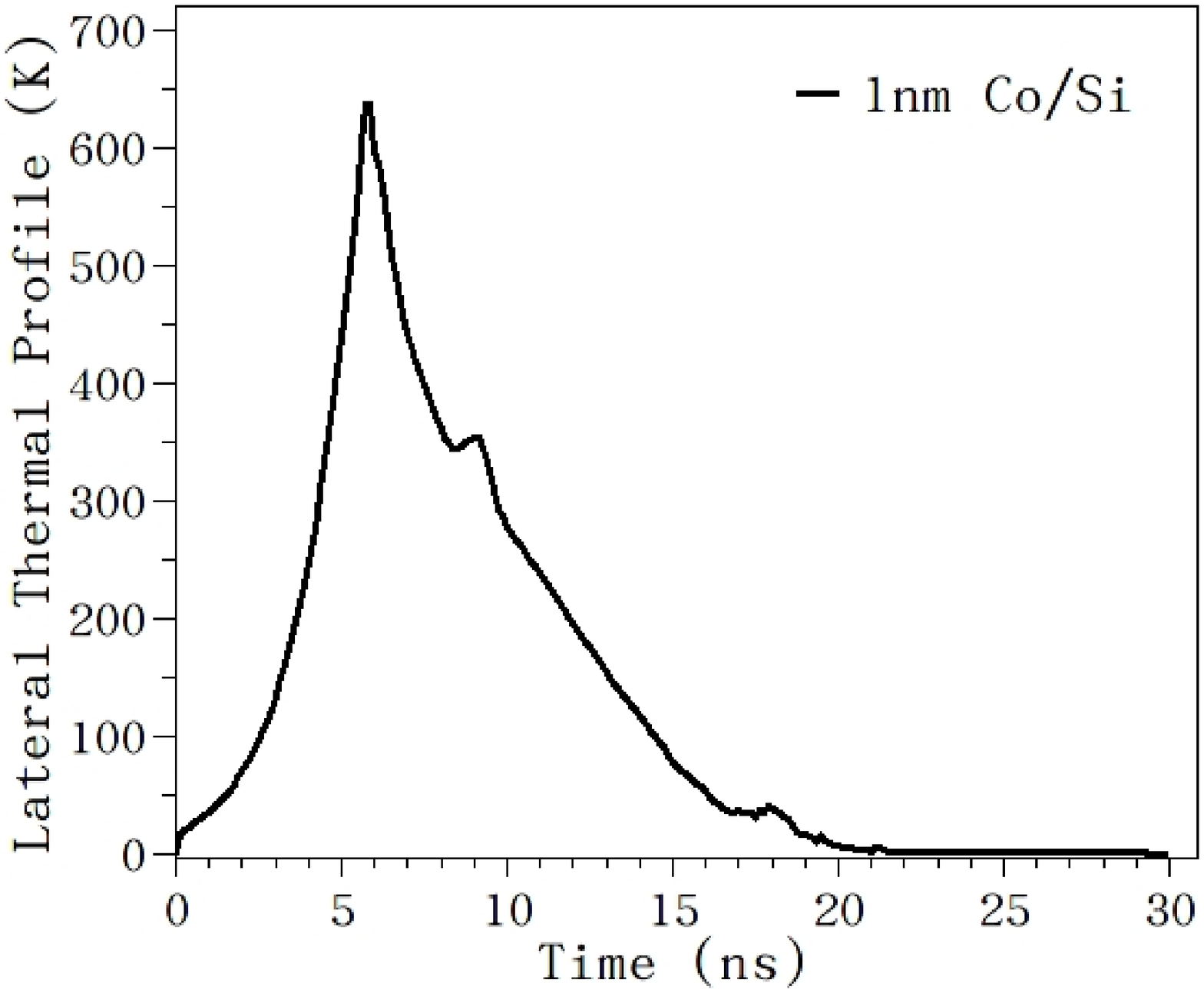}\par\end{centering}

\caption{Typical results of numerical calculations of the laser interference heating of a continuous
1 nm Co film on Si substrate. (a) The temperature profile along the surface x-direction at 10
ns showing the temperature variation is periodic with the fringe spacing of $\Lambda=400\, nm$.
(b) The time dependence of the temperature at the fringe maxima and minima. (c) The lateral thermal
profile as a function of time obtained from (b) by subtracting the temperature at any given time.
\label{fig:Typical-results-of}}
\end{figure}

\pagebreak

\begin{figure}[!tph]
\begin{centering}\subfigure[]{\includegraphics[height=3.5in,keepaspectratio]{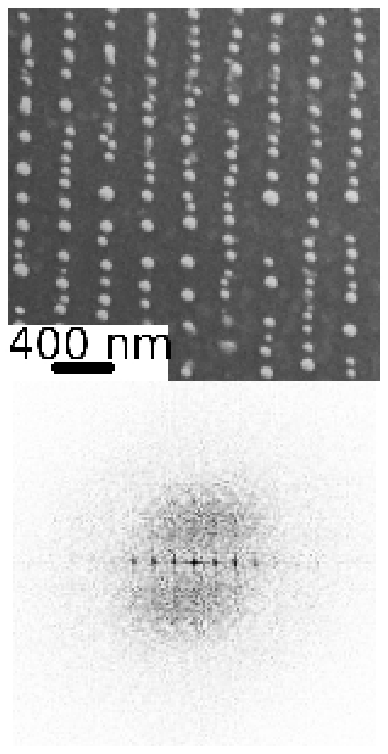}}\subfigure[]{\includegraphics[height=3.5in,keepaspectratio]{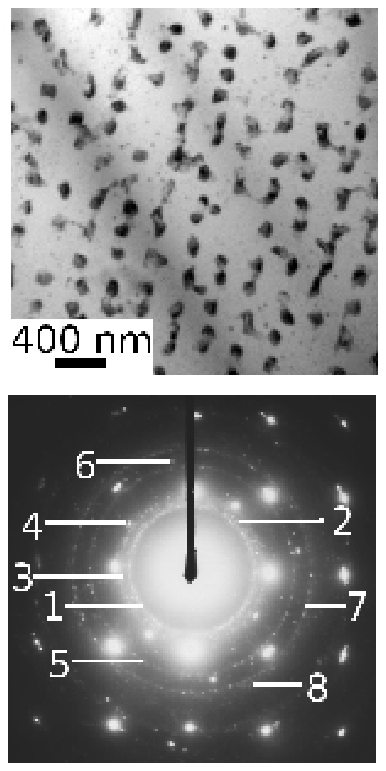}}\par\end{centering}

\caption{Spatial ordering and microstructure following dynamic patterning at energy $E_{1}<E_{Si}$.
(a) SEM image and power spectrum of the pattern in a 4 nm Co film deposited at 1 nm/min. The
nanoparticles are arranged in rows which have long range order with a periodic length scale of
400 nm, which is consistent with the interference length scale $\Lambda$. The LRO is visible
as diffraction spots in the power spectrum. No ordering is visible for the particles within each
row, as evidenced by the diffuse form of the power spectrum in directions perpendicular to the
spots\emph{.} (b) TEM micrograph and corresponding selected area diffraction pattern of a 6 nm
Co film deposited at 1 nm/min. The LRO period is consistent with the fringe spacing of 400 nm.
Indexing of the spots and rings, indicated by numbers on the diffraction image, suggest the presence
of Co metal and a small fraction of the metal-rich silicide phase Co$_{\text{3}}$Si (Table \ref{tab:PhaseInterf}).
\label{fig:Spatial-orderingInterference}}
\end{figure}

\pagebreak

\begin{figure}[!tph]
\begin{centering}\subfigure[]{\includegraphics[height=3in,keepaspectratio]{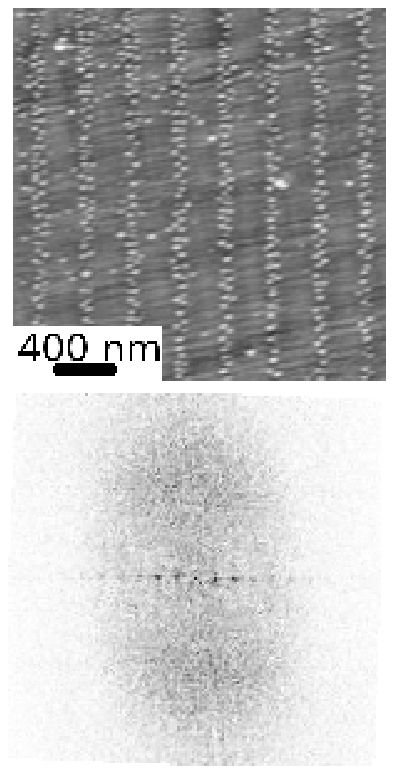}}\subfigure[]{\includegraphics[height=3in,keepaspectratio]{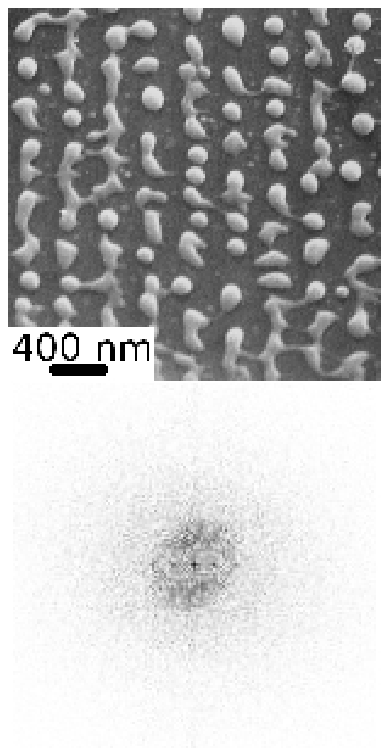}}\subfigure[]{\includegraphics[height=3in,keepaspectratio]{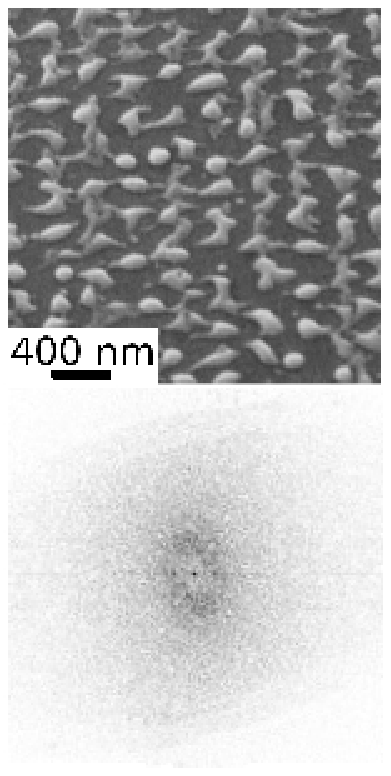}}\par\end{centering}

\caption{Pattern morphology imaged in the SEM and corresponding $PS$ following dynamic patterning in
energy regime $E_{1}$ as a function of film thickness. (a) 0.5 nm film deposited at 0.1 nm/min;
(b) 15 nm film deposited at a rate of 1 nm/min; (c) 25 nm film deposited at a rate of 1 nm/min.
In all three cases, the LRO length scale (400 nm) and wavevector direction corresponded to the
interference fringes. The average particle size increased with increasing film thickness and
therefore the nature of the ordering decreased presumably due to coalescence of the large particles
across the lines. This decrease in ordering quality is visible from the poorly defined diffraction
spots for the 25 nm film. Figures (a) and (b) are from our reference \citep{zhang05a} \label{fig:ThickDepInterf}}
\end{figure}

\pagebreak

\begin{figure}[!tph]
\begin{centering}\includegraphics[width=2in,keepaspectratio]{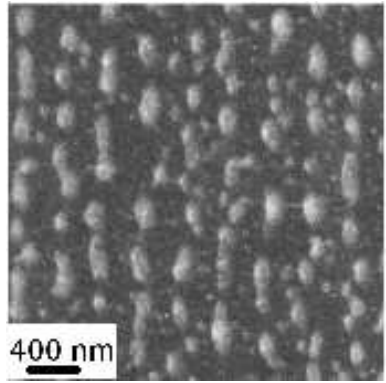}\par\end{centering}

\caption{Role of the native oxide layer. SEM micrograph of patterning done in regime $E_{1}$ for Co
on Si surface following removal of the native oxide. The morphology shows evidence for a strong
reaction between the metal and substrate and is in striking contrast to the result of similar
patterning performed on the Si surface with the native oxide layer, as shown in Fig. \ref{fig:Spatial-orderingInterference}(a).
 \label{fig:Role-of-the-oxide} }
\end{figure}

\pagebreak

\begin{figure}[!tph]
\begin{centering}\includegraphics[height=2in,keepaspectratio]{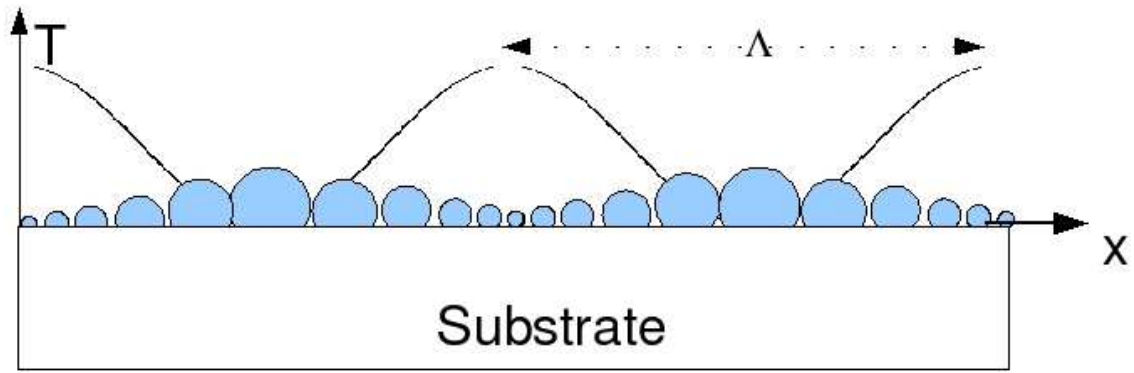}\par\end{centering}

\caption{Model for the dynamic patterning process. During early stages of deposition and simultaneous
irradiation surface diffusion results in a film morphology described by metal clusters of varying
size, with the smaller clusters in the higher temperature regions of the laser interference fringe.
The final pattern results following melting and liquid motion from the smaller clusters to the
larger ones. \label{fig:Model-for-patterning}}
\end{figure}

\begin{figure}[!tph]
\begin{centering}\subfigure[]{\includegraphics[height=3.5in,keepaspectratio]{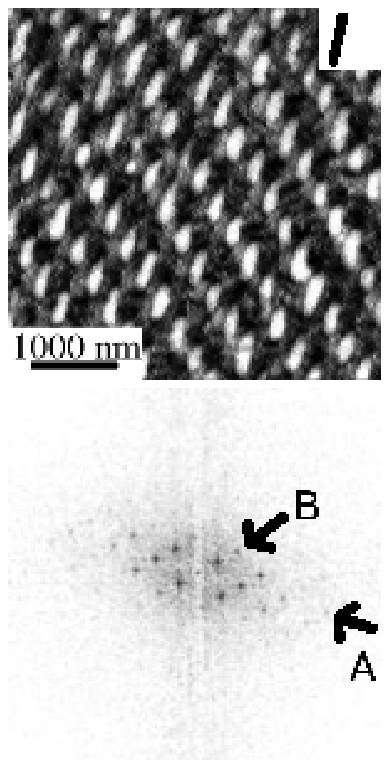}}\subfigure[]{\includegraphics[height=3.5in,keepaspectratio]{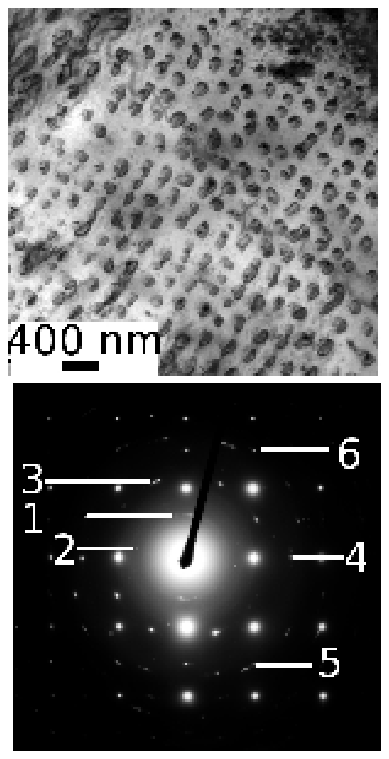}}\par\end{centering}

\caption{Spatial ordering and microstructure following dynamic patterning at energy $E_{2}>E_{Si}$.
(a) AFM image and power spectrum of the surface nanostructure for a 2 nm Co film deposited at
1 nm/min. The direction of the interference fringe is marked in the top right corner of the AFM
image. The surface features with periodicity in the direction consistent with the interference
fringe have a periodic spacing length scale of 270 nm, which is not consistent with the interference
length scale $\Lambda=400\, nm$. This period is visible as diffraction spots along direction
\emph{A} in the \emph{PS}. In addition, clear LRO is visible for the particles within each row
leading to a 2-dimensional periodic structure. This is also evidenced in the power spectrum,
with the additional diffraction spots visible along direction \emph{B}. (b) TEM micrograph and
corresponding selected area diffraction pattern of the dynamically patterned 2 nm Co film. The
2-D nature of the pattern is clearly visible. Indexing of the visible spots (Table \ref{tab:PhaseSEW})
indicates that the most dominant phase is the Si-rich silicide CoSi$_{\text{2}}$. \label{fig:Spatial-orderingSEW}}
\end{figure}

\begin{figure}[!tph]
\begin{centering}\subfigure[]{\includegraphics[height=3.5in,keepaspectratio]{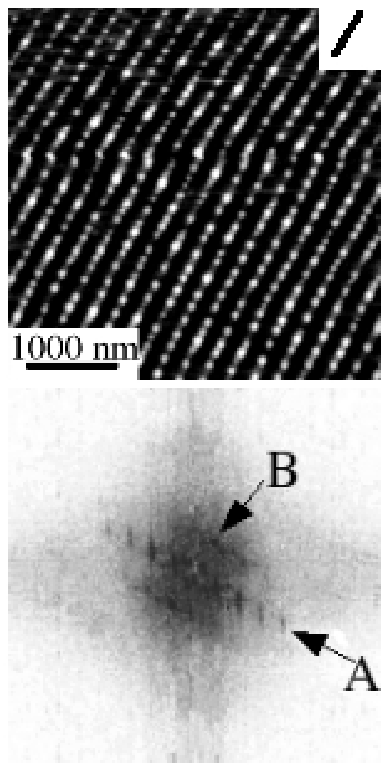}}\subfigure[]{\includegraphics[height=3.5in,keepaspectratio]{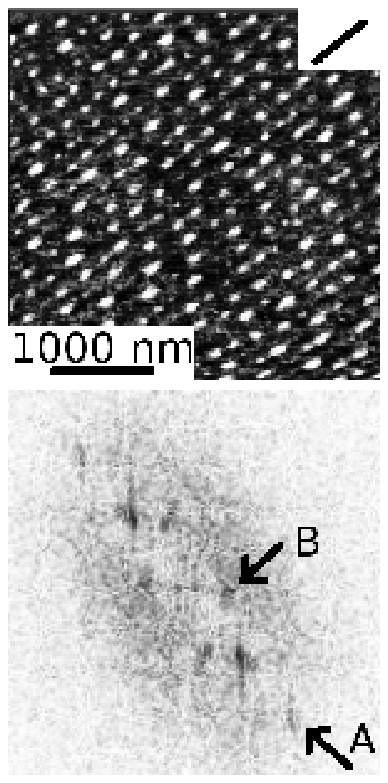}}\par\end{centering}

\caption{Pattern morphology imaged in the AFM with corresponding $PS$ following dynamic patterning in
energy regime $E_{2}$ as a function of film thickness. The line in the inset of the AFM figures
point along the expected 2-beam fringe direction. (a) 0.5 nm film deposited at 0.1 nm/min; (b)
1 nm film deposited at a rate of 1 nm/min. In both cases, a 2-D pattern is visible. Similar to
the 2 nm case (Fig. \ref{fig:Spatial-orderingSEW}) the LRO is visible as spots in the \emph{PS}
along direction \emph{A} and has spacing $L<\Lambda$ of 260 nm and 238 nm for the 0.5 and 1
nm film respectively. This length scale is similar to the 2 nm film of 249 nm. On the other hand,
the average feature size increased with increasing film thickness and the LRO indicated by spots
along \emph{B} was determined to be 172 and 474 nm respectively, as compared to 522 nm for the
2 nm film. \label{fig:ThickDepSEW}}
\end{figure}

\pagebreak

\begin{table}[!tph]
\begin{centering}\begin{tabular}{|c|c|c|c|}
\hline 
\#&
$d^{Expt}(\textrm{\AA)}$&
$d(hkl)-Co-hcp\, phase$&
$d(hkl)-Co_{3}Si$\tabularnewline
\hline
\hline 
1&
$2.41\pm0.1$&
&
2.49 (110)\tabularnewline
\hline 
2&
$2.23\pm0.1$&
2.17 (100)&
\tabularnewline
\hline 
3&
$2.05\pm0.1$&
2.02 (002)&
2.03 (002)\tabularnewline
\hline 
4&
$1.91\pm0.09$&
1.91 (101)&
1.9 (201)\tabularnewline
\hline 
5&
$1.43\pm0.07$&
1.48 (102)&
\tabularnewline
\hline 
6&
$1.33\pm0.06$&
&
1.36 (003)\tabularnewline
\hline 
7&
$1.25\pm0.06$&
1.24 (110)&
\tabularnewline
\hline 
8&
$1.16\pm0.06$&
1.15 (103)&
\tabularnewline
\hline
\end{tabular}\par\end{centering}

\caption{Indexing of diffraction spots and rings for the SAD pattern in Fig \ref{fig:Spatial-orderingInterference}(b)
obtained in regime $E_{1}$. The best matching was obtained for crystalline phases corresponding
to the bulk Co hcp phase and the metal rich hexagonal Co$_{\text{3}}$Si silicide phase. \label{tab:PhaseInterf}}
\end{table}

\begin{table}[!tph]
\begin{centering}\begin{tabular}{|c|c|c|}
\hline 
\#&
$d^{Expt}(\textrm{\AA)}$&
$d(hkl)-CoSi_{2}$\tabularnewline
\hline
\hline 
1&
3.12&
3.1 (111)\tabularnewline
\hline 
2&
2.62&
2.68 (200)\tabularnewline
\hline 
3&
1.62&
1.55 (222)\tabularnewline
\hline 
4&
1.23&
1.23 (331)\tabularnewline
\hline 
5&
1.11&
1.09 (422)\tabularnewline
\hline 
6&
1.06&
1.03 (511)\tabularnewline
\hline
\end{tabular}\par\end{centering}

\caption{Indexing of diffraction spots and rings for the SAD pattern in Fig \ref{fig:Spatial-orderingSEW}(b)
obtained in regime $E_{2}$. The best matching was obtained for crystalline phases corresponding
to the Si-rich cubic CoSi$_{\text{2}}$ silicide phase. \label{tab:PhaseSEW}}
\end{table}

\end{document}